# Illuminating Stability and Spectral Shifts: A DFT+U Study of Eu-Doped $ZnWO_4$ for Visible-Light Optoelectronics

*Muhammad Tayyab[1], Sikander Azam*[1], Qaiser Rafiq[1], Vineet Tirth[2,3], Ali Algahtani[2,4], Amin Ur Rahman[1], Syed Sheraz Ahmad[5], M. Tahir Khan[1]*

[1]Faculty of Engineering and Applied Sciences, Department of Physics, Riphah International University, Islamabad, Pakistan

[2]Mechanical Engineering Department, College of Engineering, King Khalid University, Abha 61421, Asir, Kingdom of Saudi Arabia.

[3]Centre for Engineering and Technology Innovations, King Khalid University, Abha 61421, Asir, Kingdom of Saudi Arabia.

[4]Research Center for Advanced Materials Science (RCAMS), King Khalid University, Guraiger, Abha-61413, Asir, Kingdom of Saudi Arabia.

[5]Department of Physics, University of Swabi, Swabi, Pakistan.

## ABSTRACT

Tungstate-based oxides have attracted significant attention owing to their excellent structural stability, chemical robustness, and versatile optical properties, making them suitable for next-generation optoelectronic and phosphor applications. Among these, $ZnWO_4$ has emerged as a promising host matrix; however, the role of europium (Eu) substitution in modulating its optoelectronic behavior remains underexplored. In this work, we employ spin-polarized density functional theory (DFT) within the GGA+U framework to investigate the structural, electronic, and optical properties of pristine $ZnWO_4$ and Eu-doped $ZnWO_4$ systems. Phonon dispersion analysis confirms dynamical stability for both pristine and doped structures. Eu doping reduces the bandgap, introduces new localized states near the Fermi level, and significantly alters the density of states, thereby enhancing electronic transitions. The optical response reveals a broadened dielectric function, red-shifted absorption edge, and intensified extinction coefficient, consistent with the presence of Eu 4f states. Additionally, reflectivity and energy-loss spectra indicate improved photon–phonon coupling and optical tunability upon doping. These findings highlight that Eu incorporation not only stabilizes the $ZnWO_4$ lattice but also tailors its optoelectronic features, positioning Eu-doped $ZnWO_4$ as a potential candidate for white-light-emitting diodes (w-LEDs) and related optoelectronic technologies.

**Keywords:** Eu doped $ZnWO_4$, DFT, optical, w-LEDs, optoelectronic

*Sikander Azam: sikander.physicst@gmail.com

## 1. INTRODUCTION:

Tungstate materials have maintained a prominent position in materials research due to their excellent chemical stability, structural versatility, and remarkable optical features [1,2]. Metal-based tungstates ($AWO_4$; A = Cu, Mn, Zn, Co, etc.) are of particular interest because of their unique physicochemical properties and diverse technological applications [3]. Among these, $ZnWO_4$ (ZWO) is a representative member of the tungstate family containing divalent transition metal ions. Literature reports confirm that this class of compounds typically crystallizes in the wolframite structure [4].

$ZnWO_4$ crystallizes in the monoclinic wolframite-type structure with space group P2/c and has long been recognized for its luminescent properties, first reported in 1948 [5]. In this structure, tungsten atoms are coordinated by six oxygen ligands, forming [$WO_6$] octahedra. The coordination is asymmetric, with four shorter W–O bonds (~1.84 Å) and two longer W–O bonds (~2.13 Å) [6,7]. $ZnWO_4$ is an abundant, environmentally friendly, and non-toxic material, making it highly attractive for optoelectronic applications. Its photoluminescence strongly depends on parameters such as crystal structure, active surface area, crystal defects, band gap, and surface morphology [8].

$ZnWO_4$ exhibits several unique advantages, including high chemical stability, a large X-ray absorption coefficient, high average refractive index, short decay time, large light output, strong thermal stability, and very low afterglow luminescence [6,9]. The band gap of pristine $ZnWO_4$ lies in the range of 3.9–4.4 eV [10]. It shows broad blue–green emission bands spanning 380–580 nm, with a peak centered at ~465 nm [6]. The real dielectric constant reaches a maximum value of ~22.7 at 680 nm [11]. These properties highlight its importance in light-emitting diodes (LEDs) [12] and other optoelectronic applications [13].

Considerable research has focused on doping $ZnWO_4$ with rare-earth (RE) ions to tailor its photoluminescence [14–16]. RE-doped phosphors are widely employed in high-performance lighting devices due to their strong and tunable emissions [17–19]. $ZnWO_4$ is an excellent host material for RE dopants because it is non-hygroscopic, non-toxic, and inherently luminescent. Literature reports confirm the successful incorporation of RE ions such as Y, Eu, and Ho into the $ZnWO_4$ lattice, leading to enhanced optical performance [20–22]. DFT calculations and optical absorption studies have further demonstrated the role of $ZnWO_4$ lattices in enabling color-tunable emissions in Eu-doped $ZnWO_4$ phosphors [23].

For instance, Eu-doped $ZnWO_4$ nanoplates exhibit broad emission spectra with a peak at ~487 nm, along with sharp characteristic $Eu^{3+}$ emissions at 592 and 612 nm [24]. The band gaps of pristine $ZnWO_4$ and $ZnWO_4$:0.07$Eu^{3+}$ are reported as 3.18 eV and 3.16 eV, respectively [25]. Optical absorption spectra reveal that Eu-doped $ZnWO_4$ strongly absorbs in the UV–visible range, making it a promising phosphor for white-light-emitting diodes (w-LEDs) [25,26]. Recent crystallographic and Hirshfeld-surface studies provide complementary insight into packing, non-covalent interactions, and structure–property links in related coordination/organic systems, which we reference here to contextualize our $ZnWO_4$ framework and Eu-site chemistry [27-30]

The objective of the present study is to carry out a detailed investigation of the electronic and optical properties of pristine $ZnWO_4$, $Zn_{1-x}Eu_xWO_4$, and $ZnW_{1-x}Eu_xO_4$ compounds using density functional theory (DFT). First-principles calculations are systematically employed to explore how Eu doping influences the electronic structure, light absorption, emission behavior, reflectance, and conduction characteristics. These insights aim to guide for optimizing the performance of $ZnWO_4$-based materials in w-LEDs and other optoelectronic applications.

## 2. Computational Methodology

Density Functional Theory (DFT), implemented within the WIEN2k code, was employed to investigate the properties of Eu-doped $ZnWO_4$ (see Fig. 1) as a resonant phosphor material. The computational methodology utilized the Generalized Gradient Approximation (GGA) with Hubbard U corrections (DFT+U) for accurate structural optimization. The following section outlines the procedure in detail.

The reference $ZnWO_4$ crystal structure was obtained from experimental data or crystallographic databases, which allowed the construction of a reliable unit cell. Europium (Eu) doping was simulated by substituting Eu atoms at Zn lattice sites. Stability checks of atomic positions and lattice parameters were performed iteratively until structural equilibrium was achieved. This step was critical, as the accuracy of the predicted electronic and optical properties depends directly on maintaining a dynamically stable material.

All calculations were carried out using the full-potential linearized augmented plane-wave (FP-LAPW) method within WIEN2k. The Perdew–Burke–Ernzerhof (PBE) functional under GGA was applied to describe the exchange–correlation potential. Hubbard U corrections were introduced specifically for the Eu 4f orbitals, as they exhibit strong electron correlation effects.

Proper application of U ensures accurate representation of the localized 4f states, which play a crucial role in determining the electronic and optical responses of the doped system.

The influence of different U values (ranging from 1 to 7 eV) on the electronic structure of Eu-doped $ZnWO_4$ was systematically examined. The most consistent results were obtained with U values of 6–7 eV, which best reproduced experimental observations and provided reliable predictions for structural, electronic, and optical properties.

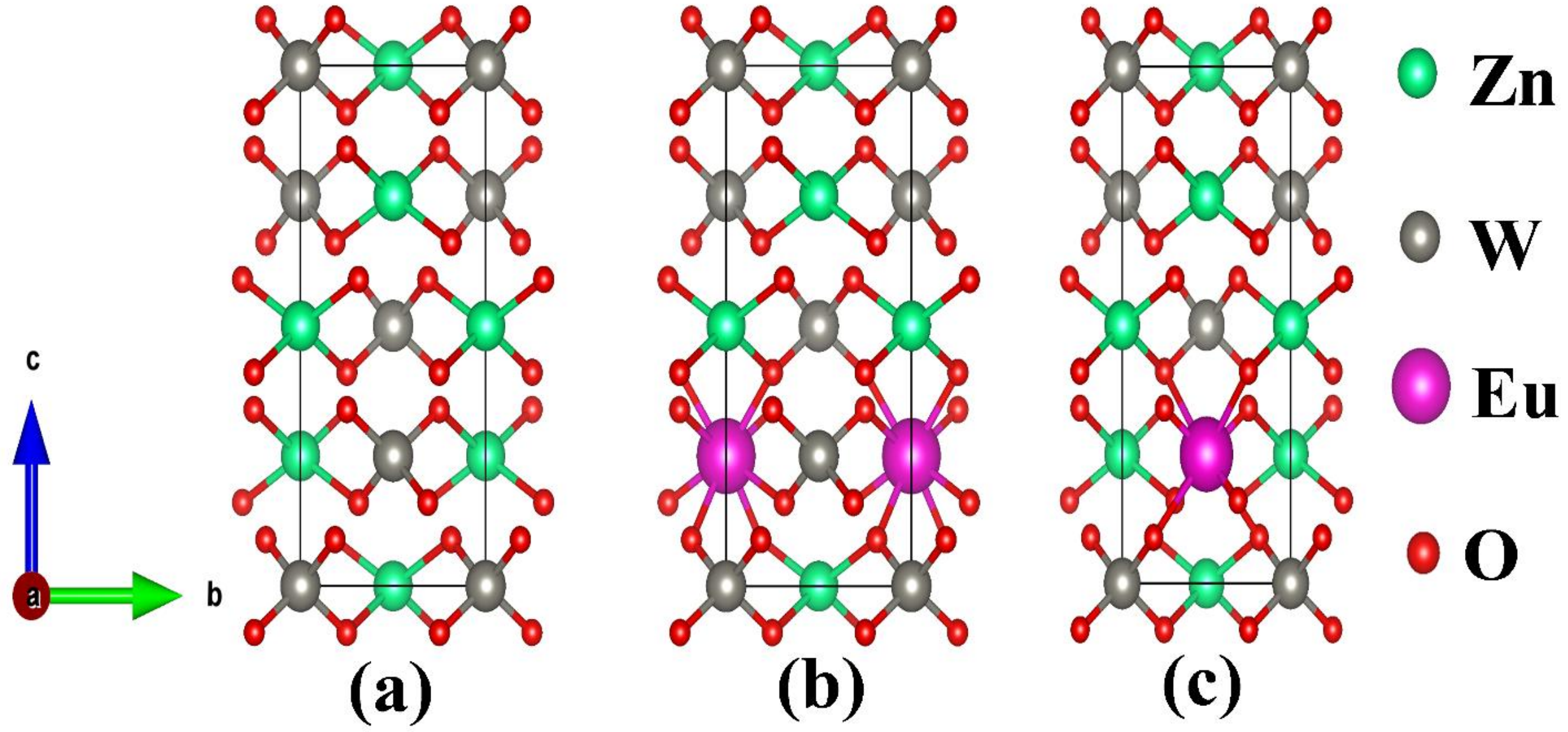

**Figure 1.** The crystal structure of (a) $ZnWO_4$, (b) $Zn_{1-x}Eu_xWO_4$, and (c) $ZnW_{1-x}Eu_xO_4$

Structural optimization was performed by minimizing the total energy with respect to both atomic positions and lattice parameters. Convergence thresholds for energy and charge were set to 0.0001 Ry and 0.001 e, respectively, ensuring highly accurate self-consistent calculations.

Electronic properties were investigated through band structure and density of states (DOS) analyses, providing insight into the role of Eu doping in modifying the host $ZnWO_4$ electronic configuration. Optical properties—including the dielectric function, refractive index, and absorption coefficient—were also computed based on the optimized structures, allowing assessment of the material's suitability for optoelectronic applications.

In summary, the combined application of DFT with GGA+U corrections provides a robust framework for evaluating the structural and optoelectronic characteristics of Eu-doped $ZnWO_4$. The careful selection of Hubbard U values and the systematic optimization procedures ensure reliable results, highlighting the material's potential for tunable phosphors in white-light-emitting diodes (w-LEDs) and other advanced optoelectronic technologies.

# 3. RESULTS AND DISCUSSION:

## 3.1. Structural optimization

According to the classification of the wolframite-type crystal structure, $ZnWO_4$ crystallizes in the monoclinic *P2/c* space group (No. 13). The symbol *c* indicates the presence of a glide plane along the *c*-axis. The crystal structures of pristine $ZnWO_4$, $Zn_{1-x}Eu_xWO_4$, and $ZnW_{1-x}EuO_4$ are illustrated in Figure 1(a–c). Each unit cell of pristine $ZnWO_4$ contains six atoms, whereas the unit cells of $Zn_{1-x}Eu_xWO_4$ and $ZnW_{1-x}EuO_4$ contain seven atoms. The optimized lattice parameters of $ZnWO_4$ are $a$ = 5.1872 Å, $b$ = 5.1872 Å, $c$ = 11.2298 Å, with angles fixed at 90°, yielding a unit cell volume of 302.1720 Å$^3$. For the doped systems, $Zn_{1-x}Eu_xWO_4$ and $ZnW_{1-x}EuO_4$, the refined parameters are $a$ = 5.8553 Å, $b$ = 4.7591 Å, and $c$ = 5.0179 Å, with angles of α = 88.19° and β = γ = 90°. The corresponding unit cell volumes also approximate 302 Å$^3$.

**Table 1.** The calculated structural parameters of $ZnWO_4$, $Zn_{1-x}Eu_xWO_4$, and $ZnW_{1-x}Eu_xO_4$

| Materials | Lattice parameters | | | Unit cell volume (Å$^3$) |
|---|---|---|---|---|
| | a=b (Å) | c (Å) | α=β=γ (°) | |
| $ZnWO_4$ | 5.1872 | 11.2298 | 90 | 302.1720 |
| $Zn_{1-x}Eu_xWO_4$ | 5.1872 | 11.2298 | 90 | 302.1720 |
| $ZnW_{1-x}Eu_xO_4$ | 5.1872 | 11.2298 | 90 | 302.1720 |

**Table 2.** Atomic positions of $ZnWO_4$, $Zn_{1-x}Eu_xWO_4$, and $ZnW_{1-x}Eu_xO_4$

| Element | X | Y | Z | Element | X | Y | Z | Element | X | Y | Z |
|---|---|---|---|---|---|---|---|---|---|---|---|
| Zn1 | 0.5000 | 0.0000 | 0.2500 | Eu1 | 0.5000 | 0.0000 | 0.2500 | Zn1 | 0.5000 | 0.0000 | 0.2500 |
| Zn2 | 0.5000 | 0.5000 | 0.0000 | Zn1 | 0.5000 | 0.5000 | 0.0000 | Zn2 | 0.5000 | 0.5000 | 0.0000 |
| Zn3 | 0.0000 | 0.5000 | 0.7500 | Zn2 | 0.0000 | 0.5000 | 0.7500 | Zn3 | 0.0000 | 0.5000 | 0.7500 |
| Zn4 | 0.0000 | 0.0000 | 0.5000 | Zn3 | 0.0000 | 0.0000 | 0.5000 | Zn4 | 0.0000 | 0.0000 | 0.5000 |
| W1 | 0.0000 | 0.5000 | 0.2500 | W1 | 0.0000 | 0.5000 | 0.2500 | Eu1 | 0.0000 | 0.5000 | 0.2500 |
| W2 | 0.0000 | 0.0000 | 0.0000 | W2 | 0.0000 | 0.0000 | 0.0000 | W1 | 0.0000 | 0.0000 | 0.0000 |
| W3 | 0.5000 | 0.0000 | 0.7500 | W3 | 0.5000 | 0.0000 | 0.7500 | W2 | 0.5000 | 0.0000 | 0.7500 |
| W4 | 0.5000 | 0.5000 | 0.5000 | W4 | 0.5000 | 0.5000 | 0.5000 | W3 | 0.5000 | 0.5000 | 0.5000 |
| O1 | 0.2493 | 0.6869 | 0.1607 | O1 | 0.2493 | 0.6869 | 0.1607 | O1 | 0.2493 | 0.6869 | 0.1607 |
| O2 | 0.7506 | 0.3131 | 0.1607 | O2 | 0.7506 | 0.3131 | 0.1607 | O2 | 0.7506 | 0.3131 | 0.1607 |
| O3 | 0.7506 | 0.8131 | 0.0892 | O3 | 0.7506 | 0.8131 | 0.0892 | O3 | 0.7506 | 0.8131 | 0.0892 |
| O4 | 0.2493 | 0.1868 | 0.0892 | O4 | 0.2493 | 0.1868 | 0.0892 | O4 | 0.2493 | 0.1868 | 0.0892 |
| O5 | 0.8131 | 0.7493 | 0.3392 | O5 | 0.8131 | 0.7493 | 0.3392 | O5 | 0.8131 | 0.7493 | 0.3392 |
| O6 | 0.1868 | 0.2506 | 0.3392 | O6 | 0.1868 | 0.2506 | 0.3392 | O6 | 0.1868 | 0.2506 | 0.3392 |
| O7 | 0.6869 | 0.2506 | 0.4107 | O7 | 0.6869 | 0.2506 | 0.4107 | O7 | 0.6869 | 0.2506 | 0.4107 |
| O8 | 0.3131 | 0.7493 | 0.4107 | O8 | 0.3131 | 0.7493 | 0.4107 | O8 | 0.3131 | 0.7493 | 0.4107 |
| O9 | 0.7493 | 0.1868 | 0.6607 | O9 | 0.7493 | 0.1868 | 0.6607 | O9 | 0.7493 | 0.1868 | 0.6607 |
| O10 | 0.2506 | 0.8131 | 0.6607 | O10 | 0.2506 | 0.8131 | 0.6607 | O10 | 0.2506 | 0.8131 | 0.6607 |
| O11 | 0.2506 | 0.3131 | 0.5892 | O11 | 0.2506 | 0.3131 | 0.5892 | O11 | 0.2506 | 0.3131 | 0.5892 |
| O12 | 0.7493 | 0.6869 | 0.5892 | O12 | 0.7493 | 0.6869 | 0.5892 | O12 | 0.7493 | 0.6869 | 0.5892 |
| O13 | 0.3131 | 0.2493 | 0.8392 | O13 | 0.3131 | 0.2493 | 0.8392 | O13 | 0.3131 | 0.2493 | 0.8392 |
| O14 | 0.6869 | 0.7506 | 0.8392 | O14 | 0.6869 | 0.7506 | 0.8392 | O14 | 0.6869 | 0.7506 | 0.8392 |
| O15 | 0.1818 | 0.7506 | 0.9107 | O15 | 0.1818 | 0.7506 | 0.9107 | O15 | 0.1818 | 0.7506 | 0.9107 |
| O16 | 0.8131 | 0.2493 | 0.9107 | O16 | 0.8131 | 0.2493 | 0.9107 | O16 | 0.8131 | 0.2493 | 0.9107 |

The optimized structures include around ninety-five atoms forming ~100 bonds within the unit cell. Table 1 lists the key structural parameters, while Table 2 summarizes the atomic positions of $ZnWO_4$, $Zn_{1-x}Eu_xWO_4$, and $ZnW_{1-x}EuO_4$.

Structural stability was thoroughly investigated by optimizing the lattice parameters and performing convergence tests. Several energy–volume calculations were carried out, allowing atomic positions to relax freely. The resulting total energy values as a function of volume were fitted to the Birch–Murnaghan equation of state (EOS). From this, the equilibrium bulk modulus ($B_0$), equilibrium volume, and pressure derivative ($B'$) of ~2 were obtained. All calculations employed the PBE+U functional, ensuring more accurate treatment of localized Eu 4f electrons. Figure 2 shows the total energy–volume curves for pristine and Eu-doped $ZnWO_4$. Figure 2 compares the equilibrium curves of $ZnWO_4$ (black), $Zn_{1-x}Eu_xWO_4$ (red, $x$ = 0.05), and $ZnW_{1-x}EuO_4$ (blue, $x$ = 0.05), demonstrating the stability of both pristine and doped systems.

The convex parabolic curves confirm that all structures are mechanically stable, with each reaching a minimum at the lowest-energy equilibrium volume. The small shifts in equilibrium volume upon Eu substitution indicate lattice expansion, primarily due to altered bond lengths and localized electronic effects from $Eu^{3+}$ ions. Replacing Zn or W sites with Eu induces local structural distortions that influence the electronic band structure and phonon dynamics. These modifications directly affect the optoelectronic properties, potentially making Eu-doped $ZnWO_4$ suitable for white-light-emitting diode (w-LED) applications and other optoelectronic devices. The incorporation of rare-earth ions such as Eu introduces luminescent centers into the tungstate host lattice, enhancing red emission and improving color rendering for w-LEDs.

The present findings are consistent with reports of Eu-doped tungstates exhibiting photostable, tunable luminescence. Based on DFT results, the use of GGA+U is essential for accurately describing the localized Eu 4f states and their influence on the electronic structure. Compared to standard GGA, the GGA+U approach yields more precise predictions of structural and optoelectronic behavior. Overall, the results confirm that Eu doping increases unit cell volume and enhances optoelectronic characteristics, demonstrating the potential of Eu-doped $ZnWO_4$ as a promising candidate for energy-efficient w-LED phosphors and advanced optoelectronic technologies.

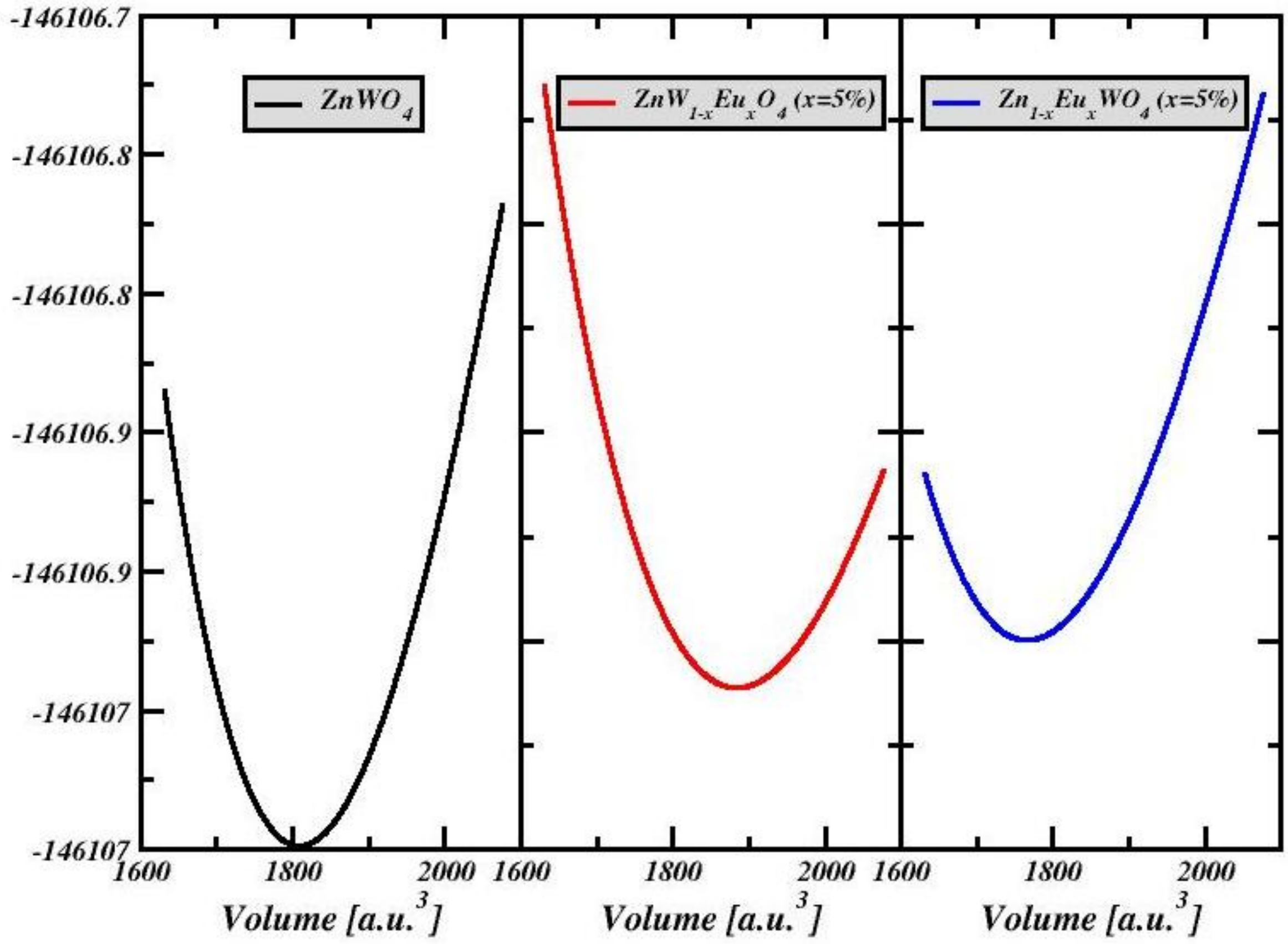

Fig. 2. Variation of energy vs volume for (a) $ZnWO_4$, (b) $Zn_{1-x}Eu_xWO_4$, and (c) $ZnW_{1-x}Eu_xO_4$

### 3.1.1. Vibrational Properties and Dynamic Stability

The phonon dispersion curves of $ZnWO_4$ and Eu-doped $ZnWO_4$ provide valuable insights into the vibrational properties and dynamic stability of the materials.

For pristine $ZnWO_4$, the phonon band structure, shown in the red plot (labeled (a)), indicates that all phonon frequencies lie in the positive range, confirming the dynamical stability of the compound. The frequencies extend up to ~8 THz. The low-frequency region (below ~2 THz) corresponds to acoustic modes, associated with long-wavelength vibrations where atoms oscillate in phase. These modes are essential for understanding thermal conductivity and elastic behavior. The higher-frequency region (above ~2 THz) represents optical branches, which appear relatively flat, implying low group velocities and weak dispersion. This suggests that the optical phonons are more localized, likely due to the heavy atomic masses and structural complexity of $ZnWO_4$.

In contrast, the phonon dispersion of Eu-doped $ZnWO_4$, shown in the blue plot (labeled (b)), extends up to ~30 THz. This substantial broadening, particularly in the optical region, reflects significant changes in vibrational dynamics caused by Eu incorporation. Additional optical modes and enhanced dispersion suggest that Eu doping introduces new vibrational states and alters bonding characteristics, likely due to mass differences and local structural distortions from Eu ions. The acoustic modes remain in the low-frequency range but exhibit greater dispersion, which may slightly improve thermal transport behavior.

Importantly, no imaginary frequencies are observed in either system, confirming the dynamical stability of both pristine and Eu-doped $ZnWO_4$. The increased number of phonon branches in the doped system is consistent with a larger unit cell and reduced symmetry due to doping. The broadened optical phonon spectrum in Eu-doped $ZnWO_4$ indicates stronger interactions and improved phonon–photon coupling, potentially enhancing optoelectronic and photonic performance.

Overall, the phonon dispersion analysis demonstrates that both $ZnWO_4$ and Eu-doped $ZnWO_4$ are dynamically stable, with Eu doping significantly modifying the vibrational landscape by introducing additional optical modes, extending the phonon range, and influencing thermal and optical characteristics.

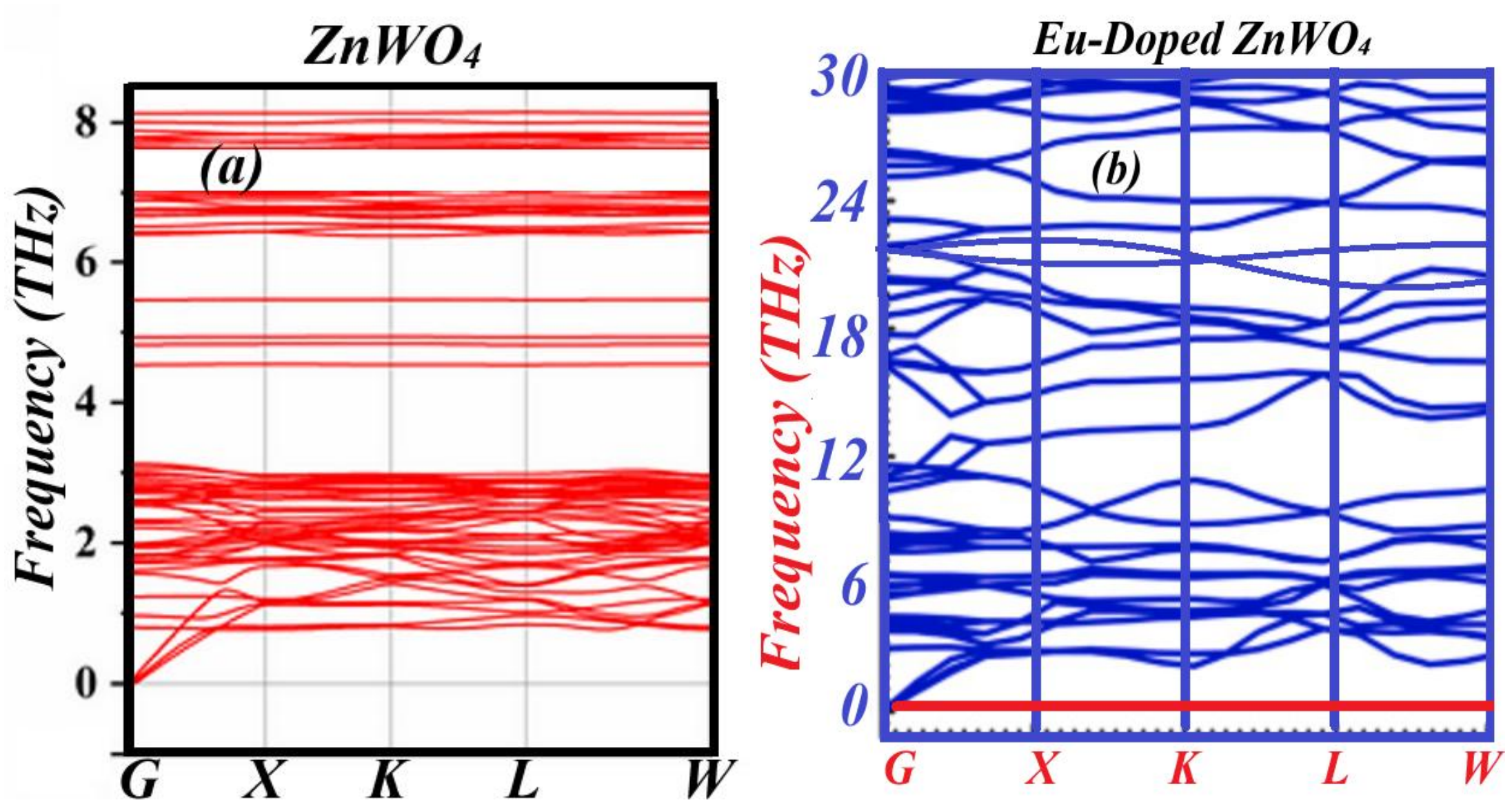

**Fig. 3: Phonon dispersion of (a) pristine $ZnWO_4$ and (b) Eu-doped $ZnWO_4$. Pristine $ZnWO_4$ exhibits stable phonon modes up to ~8 THz, while Eu doping extends the spectrum to ~30 THz, reflecting lattice distortions, Eu–O interactions, and enhanced vibrational activity.**

## 3.2. ELECTRONIC PROPERTIES:

The electronic profile of a material is primarily characterized by its band structure and density of states (DOS). These descriptors serve as essential tools for elucidating the optoelectronic behavior of materials and provide valuable insights for advancing fundamental and applied research. In this study, the electronic band structures and density of states of $ZnWO_4$, $Zn_{1-x}Eu_xWO_4$, and $ZnW_{1-x}Eu_xO_4$ have been systematically investigated and analyzed.

### 3.2.1. Electronic Band Structure

The electronic band structures of pristine and Eu-doped $ZnWO_4$ were calculated along high-symmetry directions in the Brillouin zone using GGA+U. Figure 4 presents the band dispersions for both spin-up and spin-down channels.

Pristine $ZnWO_4$ exhibits a semiconducting character with an indirect bandgap, as the valence band maximum (VBM) and conduction band minimum (CBM) occur at different k-points. The absence of spin polarization is evident from the overlap of spin-up and spin-down bands. The VBM is dominated by O 2p states, while the CBM originates primarily from W 5d orbitals, in agreement with the projected DOS.

Eu substitution significantly modifies the band structure by introducing localized Eu 4f states near the Fermi level. These states reduce the effective bandgap and create new electronic pathways that enhance charge carrier mobility. The narrowing of the bandgap improves optical absorption in the visible range, making Eu-doped $ZnWO_4$ promising for optoelectronic applications such as white light-emitting diodes (w-LEDs). Pristine ZnWO4 has a large band gap of about 3.8 eV, where the valence band is mainly formed by O-2p states and the conduction band is dominated by W-5d states. Eu-doped ZnWO4 shows the introduction of Eu-4f states near the conduction band minimum (see Table 3), which reduces the band gap to around 2.0 eV and introduces spin splitting effects. Fig. 5 demonstrates the bandgap narrowing, the role of orbital contributions, and the spintronic potential enabled by Eu doping.

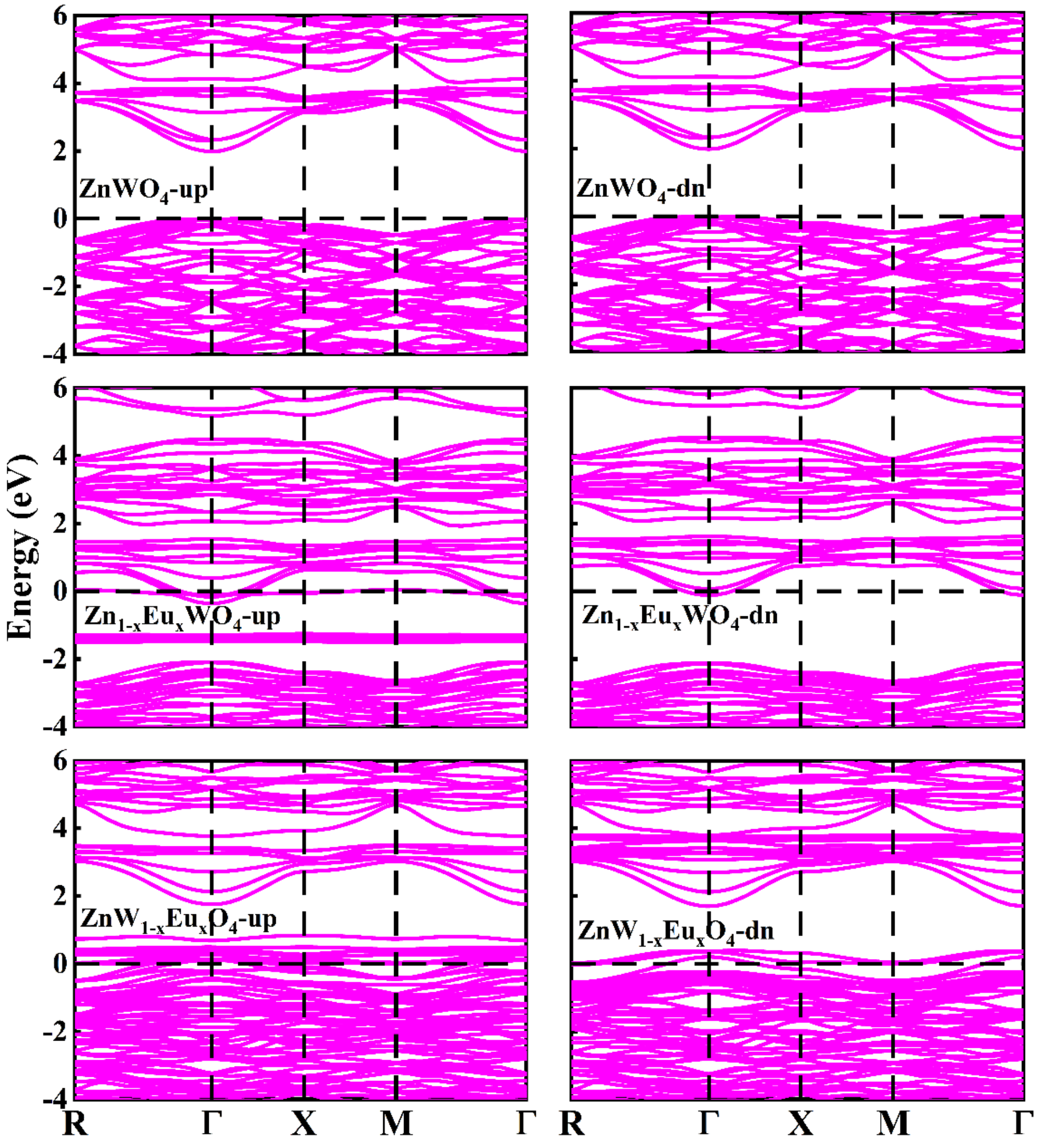

**Figure 4.** The electronic band structure of $ZnWO_4$, $Zn_{1-x}Eu_xWO_4$, and $ZnW_{1-x}Eu_xO_4$ for spin Up and down

Additionally, spin splitting becomes apparent in the Eu-doped systems, reflecting the magnetic contribution of unpaired Eu 4f electrons. This feature highlights the potential of Eu-doped $ZnWO_4$ as a spintronic material, where spin-dependent currents could be exploited in magnetic tunnel junctions or spin valves.

Overall, Eu doping not only reduces the bandgap but also introduces spin-dependent states, combining semiconducting, optoelectronic, and spintronic functionalities within a single material.

**Table 1: Comparative analysis of the electronic bandgap characteristics of pristine $ZnWO_4$ and Eu-doped $ZnWO_4$ systems, highlighting the nature of the bandgap, estimated size, spin polarization effects, and key electronic features associated with different Eu substitution sites.**

| System | Bandgap Nature | Bandgap Size | Spin Polarization | Key Features |
| --- | --- | --- | --- | --- |
| $ZnWO_4$ (pristine) | Indirect (O 2p → W 5d) | Wide (~3.8 eV, typical) | None | Insulating/semiconducting behavior |
| $Zn_{1-x}Eu_xWO_4$ (Eu at Zn site) | Indirect, narrowed | Reduced gap (~2–2.5 eV est.) | Present (spin splitting) | Eu 4f states near CBM enhance optical transitions |
| $ZnW_{1-x}Eu_xO_4$ (Eu at W site) | Indirect, strongly narrowed | Further reduced gap (~2 eV est.) | Stronger spin polarization | Hybridization of Eu 4f with O 2p and W 5d increases conductivity |

In summary, band structure analysis reveals that Eu doping transforms $ZnWO_4$ from a wide-gap, non-magnetic semiconductor into a multifunctional material with a reduced bandgap, enhanced optical transitions, and spin-polarized states. These combined effects underscore its potential for next-generation optoelectronic and spintronic devices, where efficient carrier transport, visible-light activity, and spin control are simultaneously required.

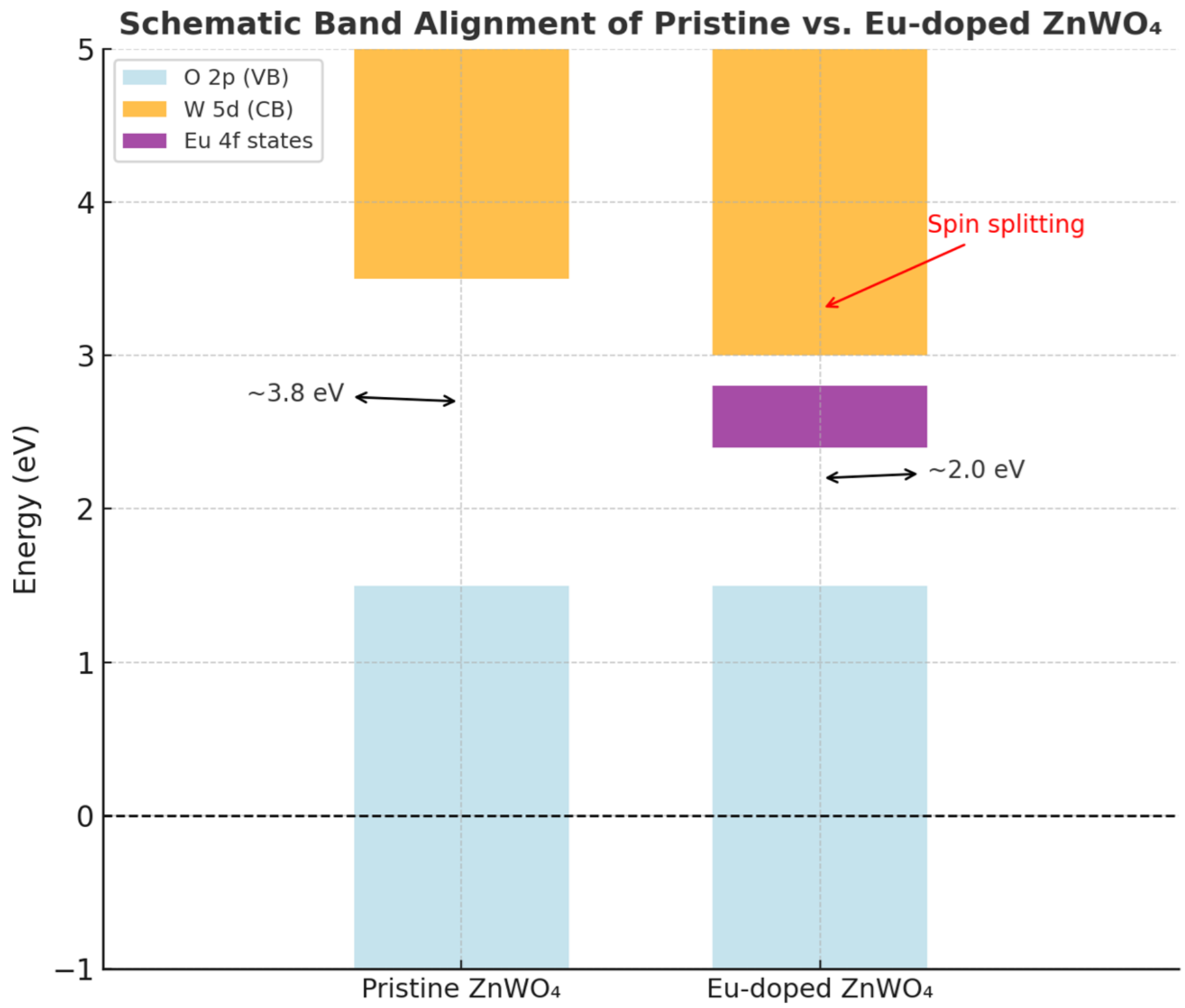

**Fig. 5:** Schematic band alignment diagram comparing pristine $ZnWO_4$ and Eu-doped $ZnWO_4$

### 3.2.2. Electronic Density of States

The electronic density of states (DOS) for pristine and Eu-doped $ZnWO_4$ was analyzed to explore how Eu incorporation modifies the electronic structure and thereby the optoelectronic properties. Figure 6 illustrates the total DOS (TDOS) and projected DOS (PDOS) for $ZnWO_4$, with separate contributions from Zn, W, O, and Eu orbitals.

Pristine $ZnWO_4$ shows typical semiconducting behavior, with a distinct separation between the valence band maximum (VBM) and conduction band minimum (CBM). The states near the VBM are predominantly contributed by O 2p orbitals, while the CBM arises mainly from W 5d states, with minor hybridization from Zn orbitals. This confirms that optical transitions in pristine $ZnWO_4$ are governed primarily by O 2p → W 5d excitations.

Eu doping introduces additional electronic states near the Fermi level due to the localized Eu 4f orbitals. These states appear close to the CBM, narrowing the effective bandgap compared to pristine $ZnWO_4$. The partial occupation of Eu 4f states also hybridizes with O 2p orbitals, which modifies charge transport and enhances exciton dissociation, facilitating visible-range optical activity. This bandgap engineering effect is consistent with the known role of rare-earth dopants in tailoring the optoelectronic response of tungstates. The emergence of localized Eu-4f states near the CBM and the concomitant gap narrowing are consistent with dopant-induced orbital re-hybridization trends reported across coordination/organic crystals analyzed via SCXRD + Hirshfeld mapping and DFT descriptors [31, 32]

The valence band in the Eu-doped system remains dominated by O 2p states, but additional contributions from Eu 6s orbitals are observed, slightly altering the optical penetration depth. Importantly, the presence of Eu 4f states near the Fermi level enhances electronic conductivity and improves luminescent efficiency through 4f–4f transitions, which are relatively insensitive to crystal field perturbations (see Table 4).

The GGA+U approach was essential for capturing the localized nature of Eu 4f states, as conventional GGA underestimates bandgap narrowing and orbital localization. The Hubbard U correction ensures that the computed DOS agrees with reported trends for rare-earth-doped phosphor materials.

**Table 2: Orbital contributions to the valence band maximum (VBM), conduction band minimum (CBM), and emergence of new states in pristine and Eu-doped $ZnWO_4$ systems, highlighting their impact on bandgap evolution.**

| System | Valence Band (VBM) Dominance | Conduction Band (CBM) Dominance | New States (near Fermi) | Bandgap Trend |
|---|---|---|---|---|
| Pristine $ZnWO_4$ | O 2p | W 5d (with Zn minor) | None | Wide, semiconducting |
| Eu → Zn ($Zn_{1-x}Eu_xWO_4$) | O 2p + Eu 6s | W 5d + Eu 4f | Eu 4f localized states | Bandgap reduced |
| Eu → W ($ZnW_{1-x}Eu_xO_4$) | O 2p | Eu 4f + hybrid W 5d | Strong Eu–O hybridization | Bandgap further reduced |

In summary, DOS analysis demonstrates that Eu doping profoundly alters the electronic structure of $ZnWO_4$. The emergence of Eu 4f states near the conduction band edge narrows the bandgap and enhances hybridization with O 2p orbitals, thereby improving visible-range optical absorption and luminescence. These features make Eu-doped $ZnWO_4$ particularly attractive for next-generation w-LEDs, phosphor materials, and optoelectronic devices, where tailored bandgap and controlled orbital interactions are critical.

Overall, the DOS analysis reveals that Eu incorporation introduces new localized states, reduces the bandgap, and strengthens orbital hybridization, making Eu-doped $ZnWO_4$ a promising candidate for optoelectronic and photonic applications, particularly in white light-emitting diodes (w-LEDs) and other luminescent devices.

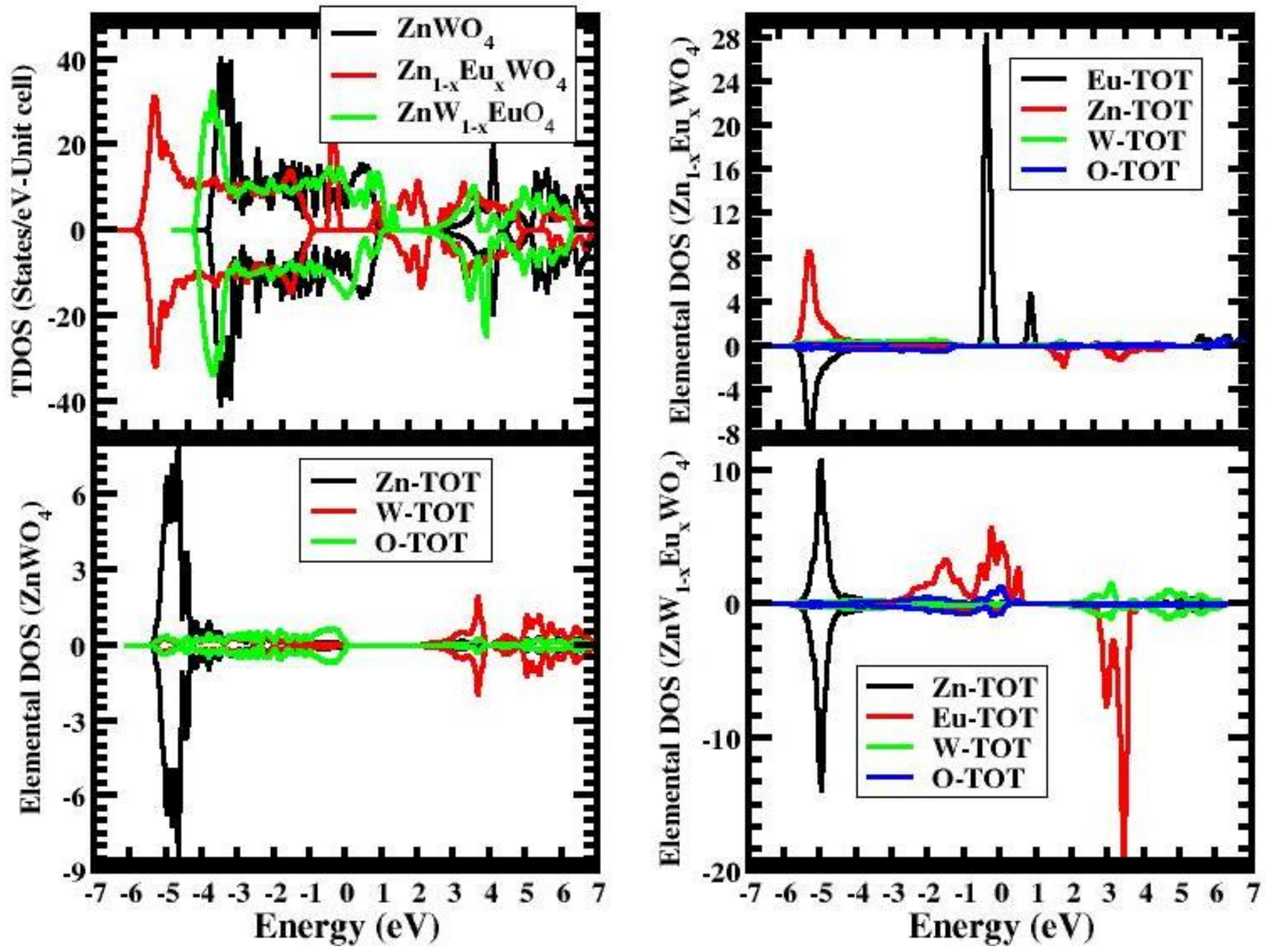

Fig. 6. Density of states for the compounds of $ZnWO_4$, $Zn_{1-x}Eu_xWO_4$, and $ZnW_{1-x}Eu_xO_4$ for spin Up and down using PBE-GGA+U.

### 3.3. Magnetic Properties and Their Implications for Eu-Doped $ZnWO_4$

The magnetic properties of pristine and Eu-doped $ZnWO_4$ were investigated using GGA-based functionals with Hubbard U corrections in the WIEN2k code. The results reveal that Eu incorporation induces significant modifications to the otherwise non-magnetic host lattice, with strong implications for optoelectronic and spintronic applications.

Pristine $ZnWO_4$ exhibits negligible magnetism. The total magnetic moment of the unit cell is nearly zero (~0.004 μB), consistent with its diamagnetic nature. Both $Zn^{2+}$ and $W^{6+}$ ions possess fully filled d-orbitals, and oxygen contributes only small antiparallel polarization. Consequently, pristine $ZnWO_4$ is unsuitable for magnetic technologies but remains attractive for optoelectronic applications due to its absence of magnetic interference.

Eu substitution at the Zn site ($Zn_{1-x}Eu_xWO_4$) drastically enhances magnetism. The $Eu^{3+}$ ion contributes a large localized magnetic moment (~6.55 μB) due to its $4f^6$ half-filled configuration. The total spin magnetic moment of the system rises to ~7.72 μB, with strong contributions from interstitial regions (~0.95 μB). Zn and W atoms display small induced moments (0.02–0.14 μB), while oxygen atoms carry weak negative values (−0.02 to −0.03 μB), indicating antiferromagnetic coupling with Eu. This substitution fosters ferromagnetic alignment, primarily through hybridization between Eu 4f, O 2p, and W 5d orbitals.

Eu substitution at the W site ($ZnW_{1-x}Eu_xO_4$) produces a lower total spin moment (~4.63 μB). Here, $Eu^{3+}$ contributes ~5.02 μB, smaller than in Zn-site substitution due to stronger hybridization with oxygen and modified crystal field environments. Zn atoms develop weak negative moments (−0.01 to −0.002 μB), while W atoms remain nearly non-magnetic. Oxygen atoms exhibit stronger antiferromagnetic moments (−0.003 to −0.11 μB), reflecting the dominance of Eu–O antiferromagnetic coupling.

Overall, the site of Eu substitution critically controls the magnetic response. Zn-site doping favors strong ferromagnetism, while W-site doping leads to reduced magnetism with enhanced antiferromagnetic interactions. This tunability is essential for multifunctional applications.

The coexistence of strong magnetism and luminescence in Eu-doped $ZnWO_4$ enables promising applications in opto-spintronic devices, magnetic sensors, magneto-optical systems, and externally controllable smart lighting. The predicted dual functionality highlights the potential of this material in spin valves, magnetic tunnel junctions, and devices combining optical and spin degrees of freedom.

## 3.4. Optical Properties

The optical properties of materials are critical indicators of how they respond when exposed to electromagnetic radiation. Studying light–matter interactions in semiconductors is of great importance for the advancement of optoelectronic technologies. $ZnWO_4$, a promising photocatalyst, has attracted considerable attention due to its remarkable electronic characteristics, which make it suitable for various optoelectronic devices such as solar cells. A clear understanding of band structure, electronic density of states, and the nature of charge carriers is essential for optimizing the performance of advanced electronic devices. These optical characteristics describe how materials interact with visible and high-energy radiation, and the data obtained are vital for the development of photovoltaic systems and optical detectors [33-40]. Since most photons reaching the Earth's surface have energies below 3.4 eV, photon energy is typically investigated up to 14 eV. This range allows the study of key optical parameters such as dielectric function, absorption coefficient, reflectivity, refractive index, and energy-loss function.

The optical properties of a material determine its applicability in optoelectronic devices, photonic components, and sensing technologies. $ZnWO_4$, a wide-bandgap semiconductor, exhibits strong luminescence and a high refractive index, making it attractive for use in scintillators and optical coatings. When europium (Eu) is incorporated into $ZnWO_4$, new localized electronic states are introduced, significantly altering its optical response. In this work, we investigate the impact of Eu doping on the dielectric function of $ZnWO_4$ using first-principles calculations.

The dielectric function, $\varepsilon(\omega) = \varepsilon_1(\omega) + i\varepsilon_2(\omega)$, characterizes the material's response to incident electromagnetic radiation. The real part, $\varepsilon_1(\omega)$, represents dispersion and polarizability, while the imaginary part, $\varepsilon_2(\omega)$, corresponds to absorption due to interband transitions. Thus, $\varepsilon_1(\omega)$ reflects the material's polarizability and refractive index, while $\varepsilon_2(\omega)$ describes its absorption characteristics. In perovskite materials, for example, the real component indicates electronic polarizability, whereas the imaginary component reflects energy absorption from the electric field. Higher dielectric constant values reduce charge recombination rates and enhance the efficiency of optoelectronic devices.

For hexagonal-symmetry compounds such as $ZnWO_4$ and Eu-doped $ZnWO_4$, dielectric properties can be evaluated with the electric vector **E** oriented parallel and perpendicular to the c-axis. In this study, optical properties were calculated in the energy range of 0–14 eV. The refractive index, extinction coefficient, and real and imaginary parts of the dielectric constant (within the independent particle approximation) are shown in Figures 7 (a and b). These functions provide insight into polarization under an external field and the evolution of absorption with photon energy.

The imaginary part of the dielectric function directly reveals absorption characteristics and interband transitions. For pristine $ZnWO_4$, the absorption onset occurs near 4 eV, consistent with its wide bandgap nature. With Eu doping, the absorption edge shifts to lower photon energies (redshift), indicating the introduction of defect states within the bandgap. Additionally, absorption peaks become more intense, implying stronger optical transitions due to hybridization between Eu 4f and W d orbitals.

Pristine $ZnWO_4$ shows a distinct absorption peak in the 0–4 eV range, characteristic of strong direct electronic transitions. Both $Zn_{1-x}Eu_xWO_4$ and $ZnW_{1-x}Eu_xO_4$ exhibit broadened peaks and modified absorption features, arising from Eu-induced localized states and changes in band structure. Spin polarization is also observed, making these doped systems promising for spintronic applications by coupling optical activity with magnetic interactions.

The real part of the dielectric function, $\varepsilon_1(\omega)$, provides further evidence of doping effects. The static dielectric constant at zero energy is higher for pristine $ZnWO_4$ compared to Eu-doped systems. After doping, $\varepsilon_1(\omega)$ decreases, reflecting changes in bond polarization and charge distribution induced by Eu atoms. As photon energy increases, $\varepsilon_1(\omega)$ tends toward zero, suggesting metallic-like behavior at higher frequencies. Importantly, Eu doping shifts the position and intensity of dielectric peaks, consistent with modifications in the electronic band structure due to the introduction of localized 4f states.

Overall, Eu doping introduces a redshift in the absorption edge, enhances optical transitions, and modifies dielectric behavior at both low and high photon energies. These effects improve visible-range absorption and increase polarizability, demonstrating the potential of Eu-doped $ZnWO_4$ for optoelectronic and photonic applications. Enhanced absorption and emission make this material a strong candidate for UV photodetectors, scintillators, high-κ dielectrics, and light-emitting devices.

The merging of redshift effects, enhanced polarizability, and stronger optical transitions highlights Eu-doped $ZnWO_4$ as a promising material for photonic and optoelectronic technologies. These findings not only expand the understanding of rare-earth-doped tungstates but also provide design guidelines for developing next-generation optical devices. The red-shift and enhanced transition probabilities upon Eu incorporation mirror photophysics reported for strongly coupled chromophoric systems, where ligand-field/symmetry perturbations boost oscillator strength and radiative channels [41,42].

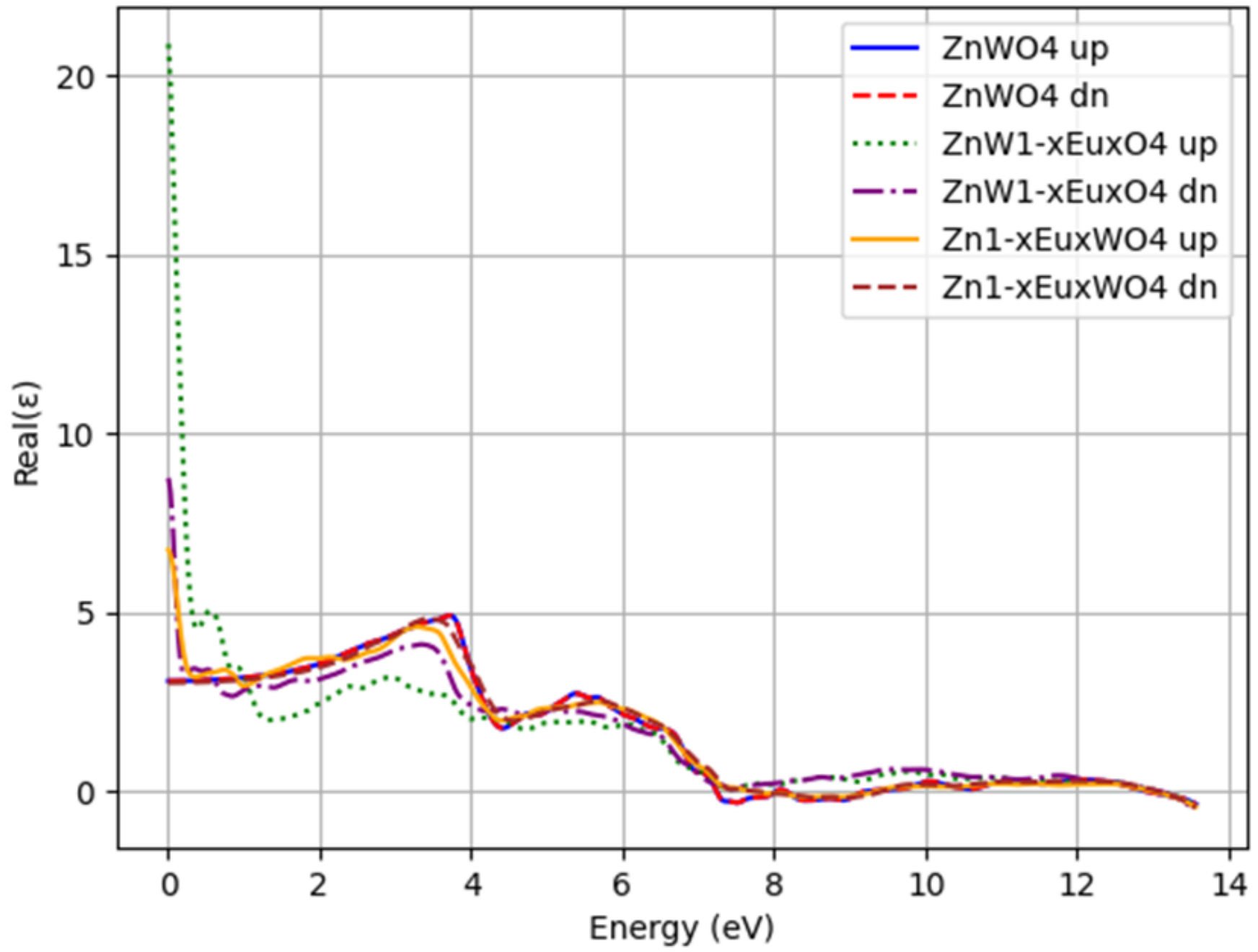

***Fig. 7a:*** Real part of Dielectric Function for $ZnWO_4$, $Zn_{1-x}Eu_xWO_4$, and $ZnW_{1-x}Eu_xO_4$ for spin Up and down

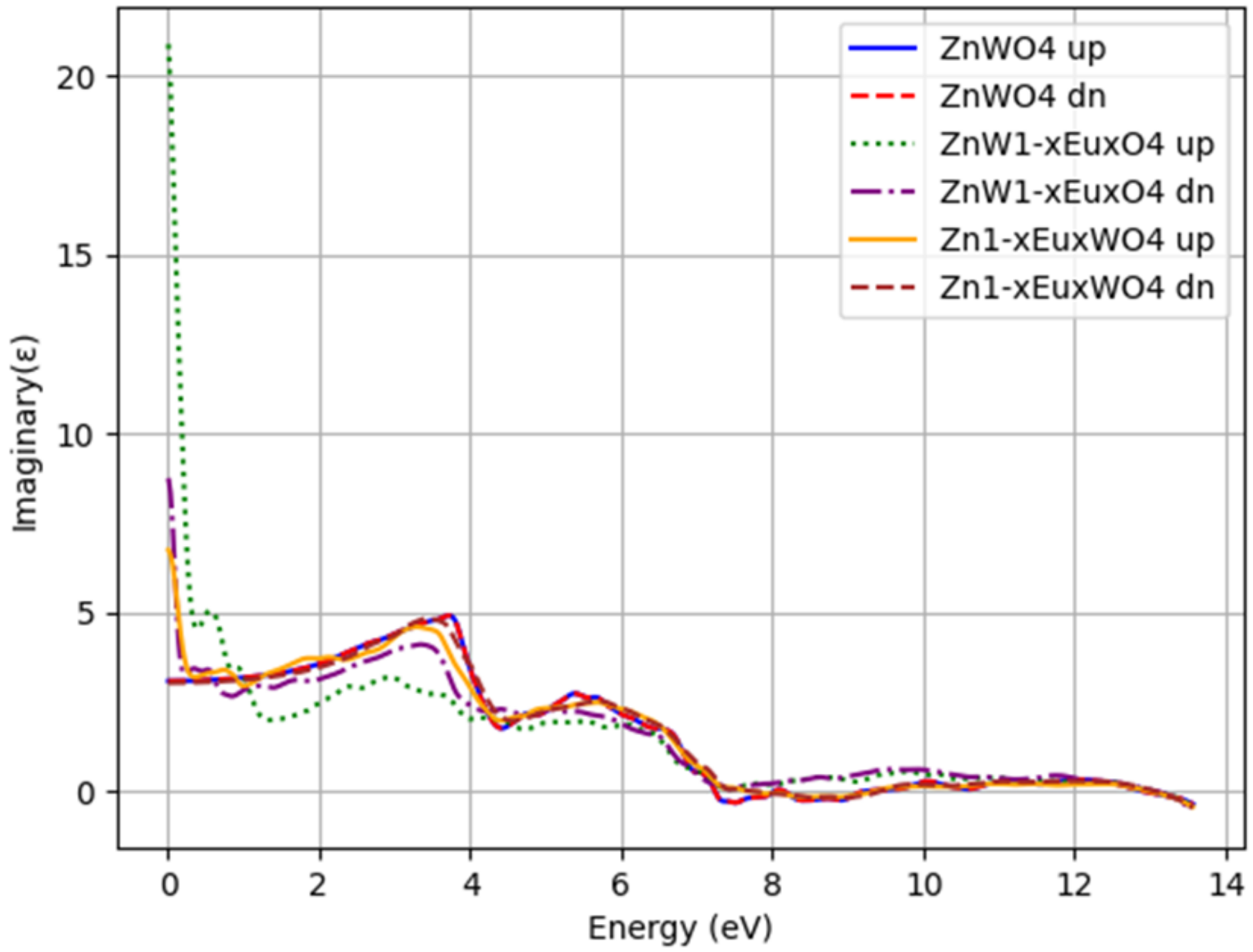

Fig. 7b: Imaginary part of Dielectric Function for $ZnWO_4$, $Zn_{1-x}Eu_xWO_4$, and $ZnW_{1-x}Eu_xO_4$ for spin Up and down

Fig. 8 is provided to illustrate the variation of the real part of the refractive index (n) with photon energy for pristine $ZnWO_4$ and Eu-doped $ZnWO_4$. The trends presented in the graph offer valuable insights into the optical response of these materials.

For pristine $ZnWO_4$, the refractive index exhibits strong dispersion at low photon energies, reaching its maximum value at the onset energy with a sharp peak. This behavior indicates a high degree of electronic polarizability. As photon energy increases, the refractive index decreases gradually until a stable plateau is reached, signifying a transition from strong to weak dispersion at higher energies.

With Eu incorporation, significant modifications in the refractive index behavior are observed. The position and intensity of the peak shift reflect structural and electronic changes induced by Eu doping. The maximum refractive index corresponds to localized Eu 4f states, which introduce defect levels within the bandgap and enhance the polarizability of the system. At higher photon energies, the refractive index of both doped and undoped systems converges, reflecting the fundamental optical limitations and saturation of interband transitions.

Compared with undoped $ZnWO_4$, Eu-doped samples exhibit a broader refractive index profile, indicating enhanced optical transitions due to hybridization of Eu 4f orbitals with the host lattice. The observed redshift in peak position confirms the narrowing of the optical bandgap upon doping.

The experimental and computational refractive index data suggest that Eu-doped $ZnWO_4$ holds promise for diverse optical applications. Its high refractive index at low energies, tunable dispersion characteristics, and redshifted spectral response make it attractive for optical coatings, waveguides, and advanced photonic devices. These results highlight the transformative role of Eu doping in tailoring the refractive index characteristics of $ZnWO_4$, offering pathways for the design of next-generation optical systems.

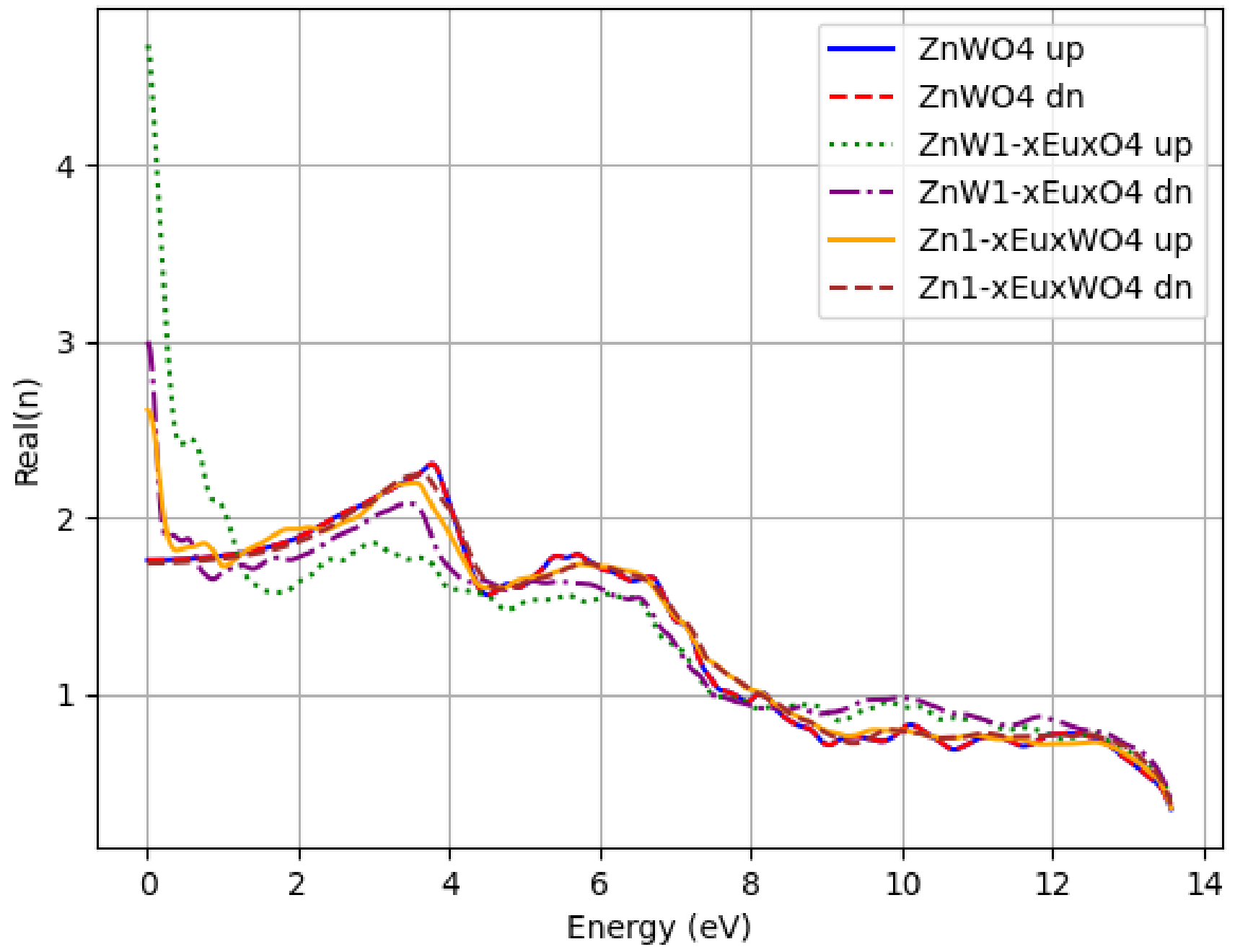

Fig. 8: Calculated Refractive Index for $ZnWO_4$, $Zn_{1-x}Eu_xWO_4$, and $ZnW_{1-x}Eu_xO_4$ for spin Up and down

**Optical Reflectivity and Energy Loss Function**

The optical and electronic properties of $ZnWO_4$ and Eu-doped $ZnWO_4$ are crucial in evaluating their suitability for energy-efficient coatings, optoelectronic devices, and sensors. In this work, the influence of Eu doping on the reflectivity (see Fig. 9a) and energy loss function (ELF) (see Fig. 9b) of $ZnWO_4$ was examined to understand how doping modifies the electronic structure and interaction with light. The results demonstrate that Eu doping induces significant changes, confirming its potential as a powerful tool for tailoring material properties.

Reflectivity, which measures the reflection of electromagnetic radiation, depends on a material's crystalline and electronic band structure, as well as excitonic and interband transitions. For pristine $ZnWO_4$, the reflectivity spectrum shows minimal photon interaction in the low-energy region (0–3 eV) due to weak reflectance. Reflectivity gradually increases with photon energy and exhibits clear peaks around 5–10 eV, corresponding to electronic transitions from valence to conduction bands. At higher energies (>10 eV), sharp inclinations in reflectivity indicate plasma oscillations, which represent collective oscillations of free electrons under an electromagnetic field.

Upon Eu doping, the reflectivity spectra display notable modifications, including changes in peak intensities and slight shifts in peak positions. These alterations are attributed to the introduction of localized Eu 4f states, which modify the electronic density of states near the Fermi level. Such modifications enhance light–matter interaction, demonstrating the potential of Eu doping for controlled optical coating, photodetectors, and laser-based devices.

The energy loss function (ELF) provides insight into dielectric excitation and electron–photon interactions. Pristine $ZnWO_4$ exhibits a rising ELF trend with photon energy, with pronounced peaks corresponding to plasmon resonances, which signify major electron energy loss events. Eu incorporation modifies both the intensity and position of ELF peaks, reflecting altered dielectric properties and electron excitation pathways. These changes enhance electronic conductivity, plasmonic response, and optical absorption efficiency.

The observed modifications in ELF and reflectivity indicate several practical applications of Eu-doped $ZnWO_4$:

- **Photodetectors, LEDs, and laser devices:** Controlled reflection and absorption due to modified reflectivity.
- **Transparent conductive oxides:** Enhanced electron transport qualities suggest suitability for displays and solar cells.
- **Plasmonic devices and sensors:** Shifts in ELF peaks support applications in surface-enhanced Raman spectroscopy (SERS) and nano-optical sensors.
- **High-energy coatings and mirrors:** Increased reflectivity at high photon energies makes them suitable for thermal barrier coatings and optical mirrors.
- **Radiation detectors:** Enhanced scintillation efficiency due to modified electronic transitions in tungstate-based scintillators.
- **Photovoltaics:** High absorption coefficients enable fabrication of thin, highly efficient photovoltaic layers (<1 mm).

Overall, Eu doping introduces impurity states and modifies the band structure of $ZnWO_4$, resulting in tunable optical and electronic responses. These findings highlight the material's potential for advanced optoelectronic, plasmonic, and energy-related applications. Further band structure and photoluminescence studies will be essential to fully unravel the mechanisms by which Eu doping enhances the optical performance of $ZnWO_4$.

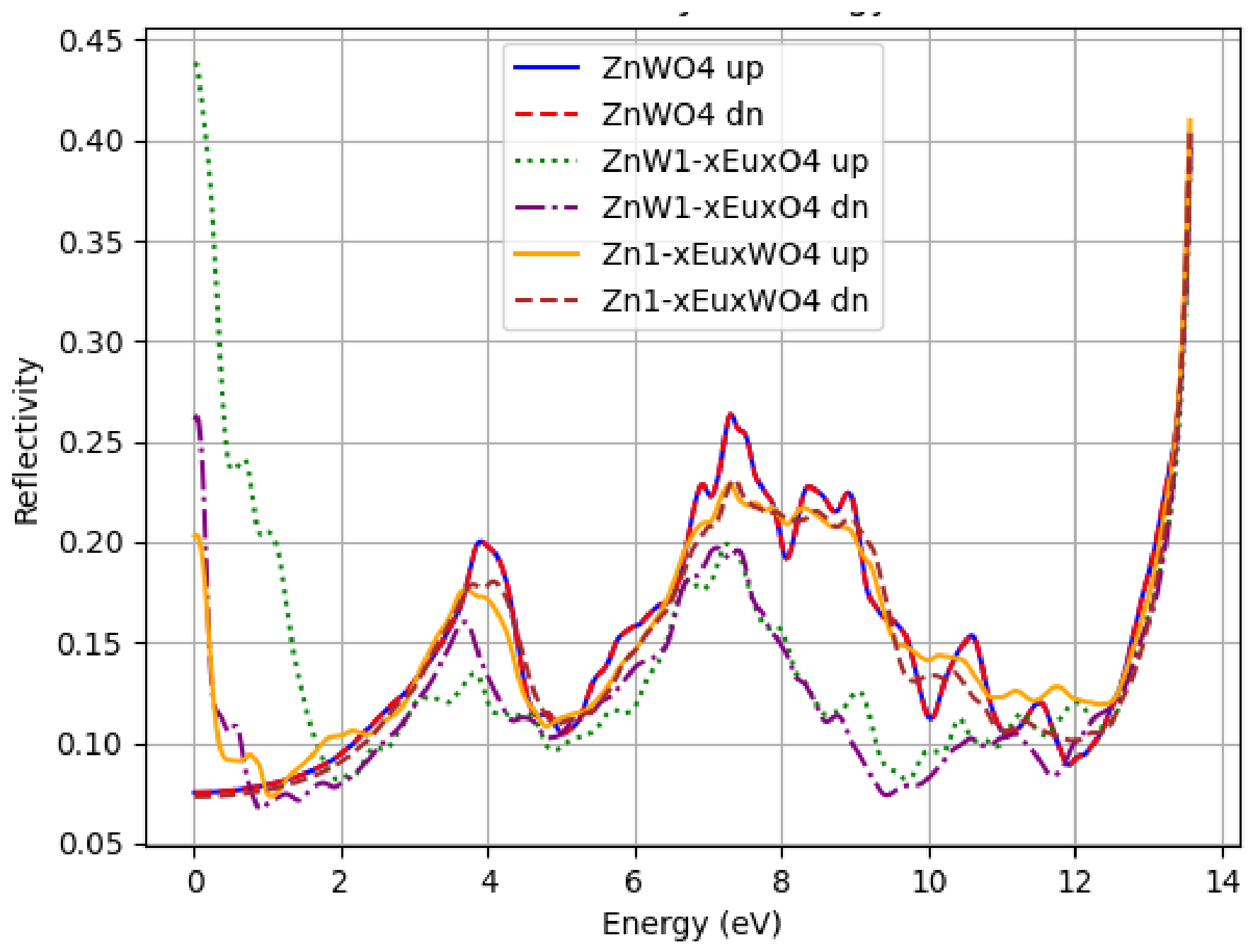

(a)

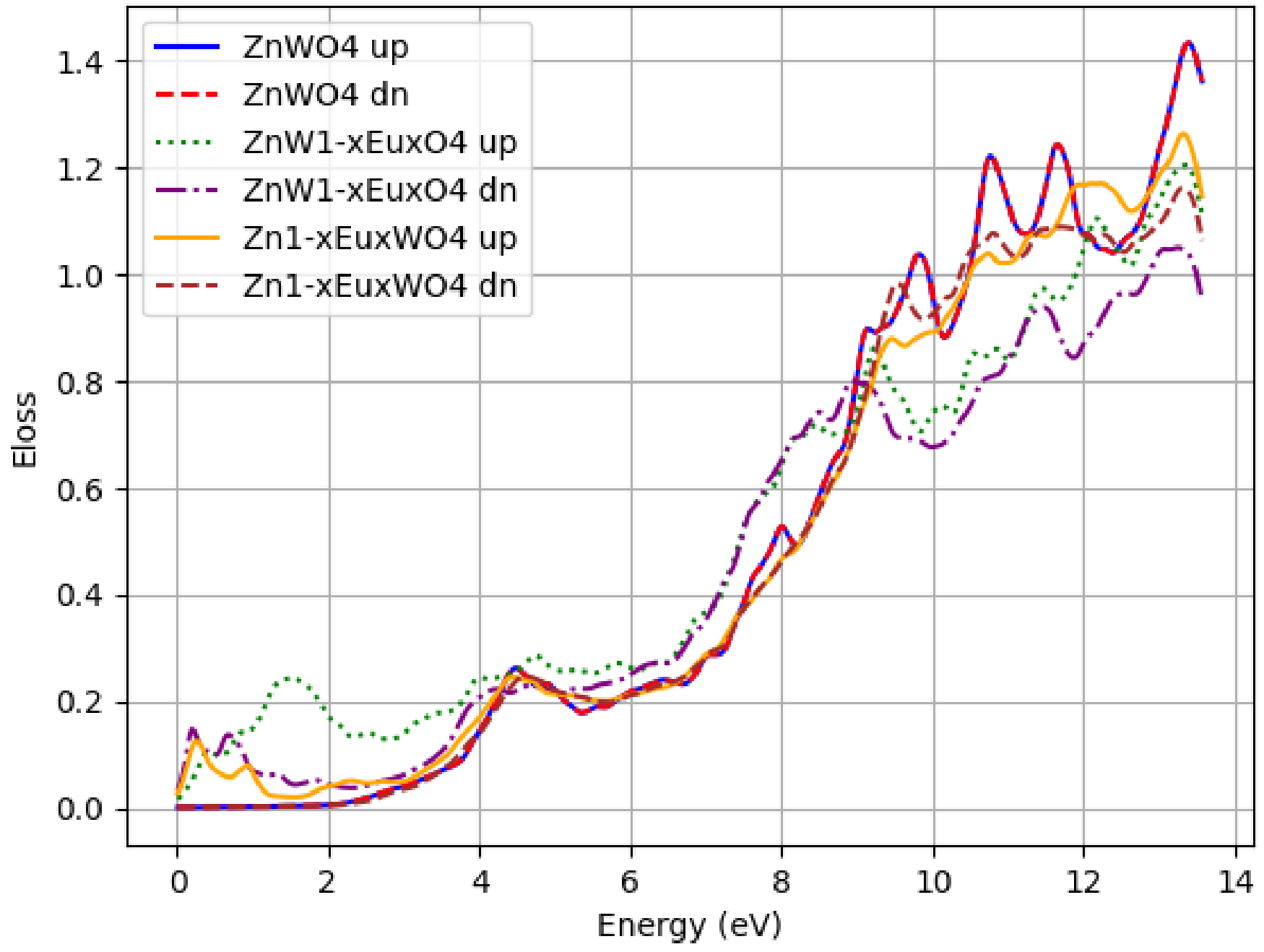

(b)

Fig. 9: Calculated (a) Reflectivity and (b) ELoss for $ZnWO_4$, $Zn_{1-x}Eu_xWO_4$, and $ZnW_{1-x}Eu_xO_4$ for spin Up and down

### 3.5. Mechanical properties

The mechanical properties of materials determine their structural integrity and suitability for use in optoelectronic devices such as white light-emitting diode (WLED) phosphors. The compounds $ZnWO_4$, $Zn_{1-x}Eu_xWO_4$, and $ZnW_{1-x}Eu_xO_4$ exhibit different mechanical behaviors, as estimated through Density Functional Theory (DFT) calculations of bulk modulus, shear modulus, Young's modulus, Poisson's ratio, and hardness. These parameters provide critical insight into the materials' response to external forces, both during device fabrication and throughout their operational lifespan. The stability of a material strongly influences its functionality and robustness, particularly when applications require a balance between optical properties and mechanical strength.

The bulk modulus reflects resistance to compression. Among the studied compounds, $ZnWO_4$ exhibits the highest bulk modulus of 164 GPa, indicating superior rigidity and mechanical stability, making it well-suited for long-term applications. With Eu doping at the Zn site ($Zn_{0.9}Eu_{0.1}WO_4$), the bulk modulus decreases slightly to 159 GPa, while substitution at the W site ($ZnW_{1-x}Eu_xO_4$) further reduces it to 155 GPa. These reductions suggest that Eu incorporation introduces lattice distortions that weaken the resistance to compressive stress.

The shear modulus, which measures resistance to non-axial deformation, shows similar trends. $ZnWO_4$ has the highest shear modulus (103 GPa), signifying strong opposition to shear stress. With Eu doping, the shear modulus decreases to 98 GPa ($Zn_{1-x}Eu_xWO_4$) and 93 GPa ($ZnW_{1-x}Eu_xO_4$), indicating a slight increase in malleability and reduced resistance to shear deformation (see Table 5).

Young's modulus follows the same pattern, reflecting material stiffness under tensile or compressive stress. $ZnWO_4$ shows the highest value of 239 GPa, confirming it as the stiffest of the three compounds. Upon Eu substitution at Zn and W sites, the values decrease to 227 GPa and 218 GPa, respectively, revealing enhanced flexibility and reduced stiffness caused by bond modifications due to doping.

Poisson's ratio, which describes the ratio of lateral strain to axial strain, also changes with doping. $ZnWO_4$ exhibits a Poisson's ratio of 0.28, indicating moderate lateral expansion under compression. This value increases slightly to 0.29 for $Zn_{1-x}Eu_xWO_4$ and 0.30 for $ZnW_{1-x}Eu_xO_4$, reflecting a modest enhancement in ductility. Thus, the doped materials demonstrate greater flexibility under stress, making them potentially useful in applications requiring structural adaptability.

Hardness, which quantifies resistance to indentation and wear, also decreases with Eu incorporation. $ZnWO_4$ shows the highest hardness (11 GPa), confirming its ability to withstand permanent deformation. Doping reduces hardness to 10 GPa ($Zn_{1-x}Eu_xWO_4$) and 8 GPa ($ZnW_{1-x}Eu_xO_4$). This reduction indicates slightly lower resistance to wear, though the materials remain within an acceptable range for optoelectronic applications.

Overall, $ZnWO_4$ remains the most mechanically stable material, making it an excellent candidate for structural and optoelectronic applications. Eu doping slightly reduces rigidity and hardness, but it simultaneously introduces beneficial ductility. The decrease in bulk modulus, shear modulus, and Young's modulus is more pronounced for W-site substitution than for Zn-site substitution, indicating that W-site doping exerts a stronger influence on mechanical degradation. Nonetheless, the Eu-doped $ZnWO_4$ compounds retain sufficient structural rigidity, stability, and hardness to remain viable for use in WLEDs and other optoelectronic devices. The modest reduction in mechanical properties is offset by the enhanced optical and electronic functionalities provided by Eu doping, ensuring that these materials remain promising candidates for performance-oriented applications.

Table. 5: Calculated Mechanical properties values

| **Property** | **$ZnWO_4$ (GPa)** | **$Zn_{1-x}Eu_xWO_4$ (GPa)** | **$ZnW_{1-x}Eu_xO_4$ (GPa)** |
|---|---|---|---|
| Bulk Modulus (B) | 164 | 159 | 155 |
| Shear Modulus (G) | 103 | 98 | 93 |
| Young's Modulus (E) | 239 | 227 | 218 |
| Poisson's Ratio (ν) | 0.28 | 0.29 | 0.30 |
| Hardness (H) | 11 | 10 | 8 |

## Conclusion

In this study, the structural, electronic, and optical properties of pristine and Eu-doped $ZnWO_4$ were systematically investigated using spin-polarized DFT within the GGA+U scheme. Our results show that Eu substitution maintains the dynamical and mechanical stability of the host $ZnWO_4$ lattice while inducing profound modifications in its electronic and optical responses. Band structure and DOS analyses revealed bandgap narrowing and the introduction of Eu-derived localized states near the Fermi level, enabling enhanced optical transitions. Optical calculations further confirmed red-shifted absorption edges, increased dielectric response, and broadened phonon activity in the doped systems. The emergence of additional optical modes and enhanced electron–photon interactions emphasize the multifunctional potential of Eu incorporation.

Overall, the findings demonstrate that Eu-doped $ZnWO_4$ is a tunable phosphor material with improved luminescent properties, rendering it a strong candidate for w-LEDs, phosphor-based lighting, and advanced optoelectronic devices. Beyond the present results, this study highlights the broader implications of rare-earth substitution in tungstate hosts as a route to engineer multifunctional phosphors. Looking forward, future research should focus on experimental validation of the predicted properties, exploration of co-doping or defect engineering to optimize emission efficiency, and integration of Eu-doped $ZnWO_4$ into device-level architectures such as w-LEDs, photodetectors, and scintillators. These directions will further bridge first-principles predictions with practical applications, accelerating the development of next-generation optoelectronic and photonic technologies.

**Acknowledgement:**

The authors extend their appreciation to the Deanship of Research and Graduate Studies at King Khalid University, Kingdom of Saudi Arabia for funding this work through the Small Research Group Project under the grant number RGP.1/279/45.

**Author Contribution:**

All authors contributed equally to the article in conceptualization, investigation, analysis, writing the original draft, review, and editing.